\pdfoutput=1

\documentclass[11pt]{article}

\usepackage{ACL2023}

\usepackage{times}
\usepackage{latexsym}

\usepackage[T1]{fontenc}

\usepackage[utf8]{inputenc}

\usepackage{microtype}

\usepackage{inconsolata}

\usepackage{latexsym}
\usepackage{graphicx}
\usepackage{amsmath}
\usepackage{algpseudocode}
\usepackage{algorithm}

\newcommand\pgpt{\texttt{parseGPT}}
\newcommand\mgpt{\texttt{matchGPT}}

\title{Interactive Learning of Hierarchical Tasks From Dialog With GPT}

\author{Lane Lawley \and Christopher J. MacLellan \\
  Georgia Institute of Technology \\
  School of Interactive Computing \\
  \texttt{\{lanetrain,cmaclell\}@gatech.edu} \\}

\begin{document}
\makeatletter
\let\@oldmaketitle\@maketitle
\renewcommand{\@maketitle}{\@oldmaketitle
  \begin{center}
  \captionsetup{type=figure}
  \includegraphics[width=0.7\textwidth]
    {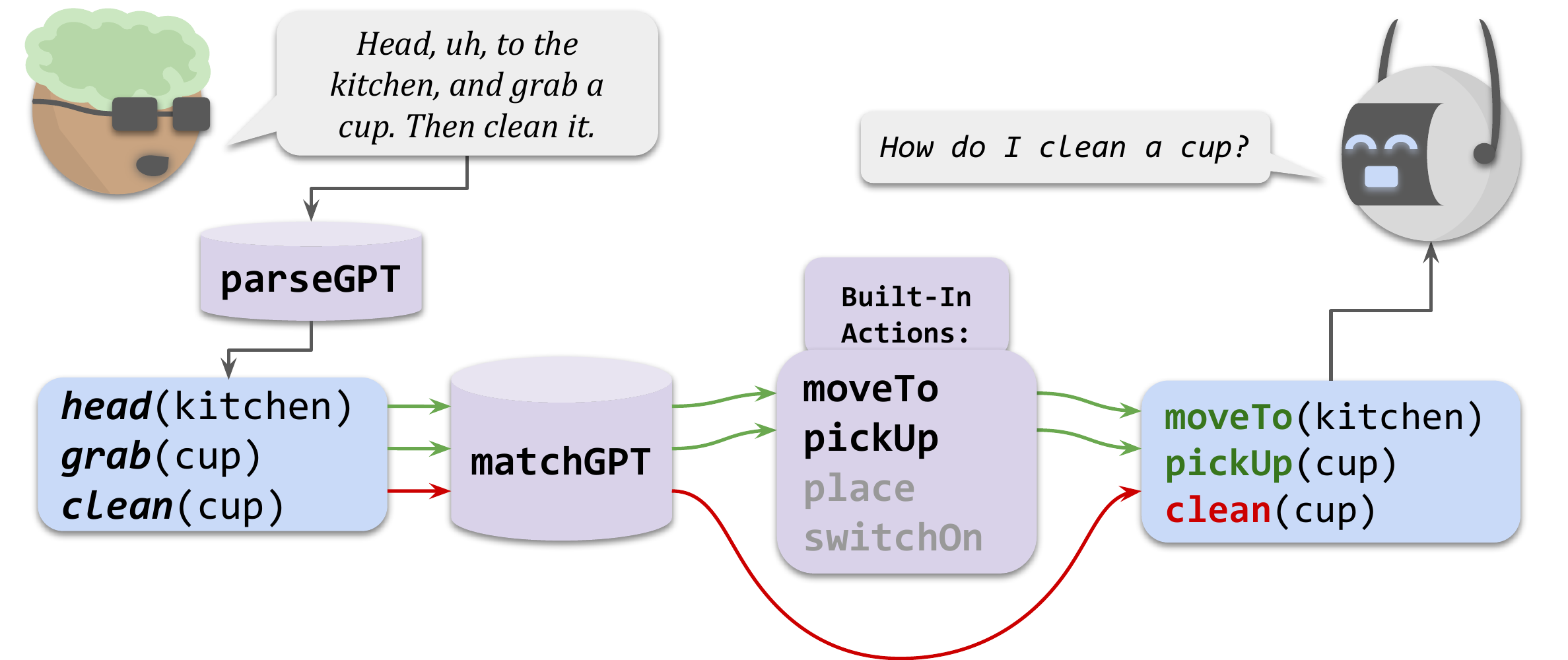}
    \caption{One iteration of a hierarchical task learning dialog in our system.}
    \label{fig:arch}
    \end{center}\bigskip}
\makeatother
\maketitle
\begin{abstract}
We present a system for interpretable, symbolic, interactive task learning from dialog using a GPT model as a conversational front-end. The learned tasks are represented as hierarchical decompositions of predicate-argument structures with scoped variable arguments. By using a GPT model to convert interactive dialog into a semantic representation, and then recursively asking for definitions of unknown steps, we show that hierarchical task knowledge can be acquired and re-used in a natural and unrestrained conversational environment. We compare our system to a similar architecture using a more conventional parser and show that our system tolerates a much wider variety of linguistic variance.
\end{abstract}

\section{Introduction}

Exciting times lie ahead in the world of task knowledge representation. The Interactive Task Learning (ITL) approach introduced by \citet{itl} has articulated a research vision dedicated to the goal of enabling machines to interactively learn general tasks in natural, human-like ways. It posits that acquired task knowledge should be general enough to be applicable in novel situations, and that it should be interpretable and modifiable, so that human teachers can actively shape the machine's understanding. ITL highlights many instructional modalities for the acquisition of task knowledge, including gestures, diagrams, and language---these modalities are also formally accounted for in the Natural Training Interaction (NTI) framework due to \citet{nti}. In this work, we focus on the language modality, endeavoring toward acquisition of human-interpetable, modular, hierarchical task knowledge from natural dialogs between humans and machines.

There has been considerable research into language-driven task acquisition. 
Early work with Instructo-Soar \citep{huffman1993} investigated how to acquire how hierarchical task descriptions for use in the Soar cognitive architecture \citep{soar}.
A decade later, this approach has evolved into Rosie \citep{rosie}, a system that can learn game and robotics task knowledge from language instruction. \citet{forbus} ground task learning in a digital sketching environment, supporting multi-modal task learning from both dialog and sketches. The PLOW system \citep{plow} integrates a demonstrative modality and a rich internal knowledge base with instructional dialogs to facilitate one-session task instruction. And, most closely related to this work, \citet{suddrey} introduce a system for learning hierarchical task representations through the recursive clarification of unknown predicates in instructional dialogs.

\begin{figure}
    \centering
    \includegraphics[width=\columnwidth]{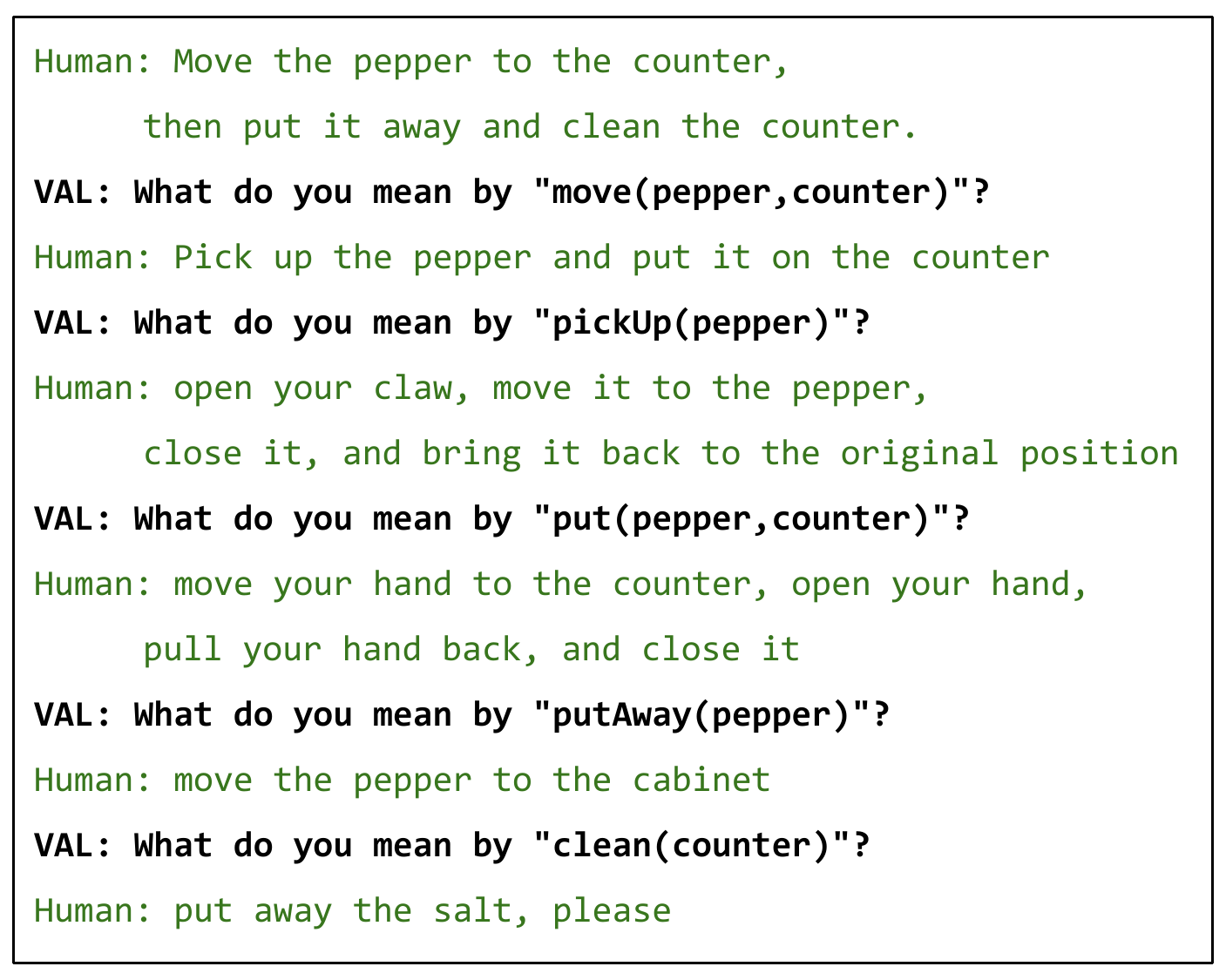}
    \caption{An example of an instructional dialog between a human instructor and our system. This example showcases GPT's paraphrase and anaphora resolution capabilities, as discussed in Section~\ref{sec:use_of_gpt}.}
    \label{fig:val_dialog}
\end{figure}

Each of these systems, however, suffers the characteristic brittleness of the classical syntactic and semantic parsers they invariably use to convert natural instructional dialogs into task knowledge. Syntactic forms vastly outnumber their semantic counterparts, and problems like anaphora resolution, paraphrasing, and grammatical and transcription errors frequently take the ``natural'' out of ``natural language'' as the first casualty of their mitigation. Each of the systems described above not only falls subject to myriad parser errors on malformed input, but also lacks the ability to perform adequate paraphrase resolution when parsed verb structures do not match known ones exactly, generally relying on manually constructed knowledge bases of synonyms to solve even a portion of this problem.
These issues make it unlikely that these systems will successfully achieve the ITL vision with real users--currently they only work with users in the lab that know their idiosyncratic syntax. 

Here, we present an alternate strategy for mitigating the sheer variety of natural language: by exploiting the virtual mastery of the \textit{form} of natural English apparently acquired by large language models \citep{bert-rediscovers-nlp}, we demonstrate the acquisition of task knowledge in a far less restrictive, far more natural dialog setting than possible in prior work. We integrate a GPT-family language model \citep{gpt-3} in a careful and principled way, using it to map natural dialog into a symbolic domain. The GPT model is used for two sub-tasks: the semantic parsing of text into predicate-argument structures, and the semantic unification of those structures to already-known actions with the same meaning. The former task allows the system to deal with error-laden and grammatically unrestricted input dialog, and the latter allows it to perform paraphrasal mappings that take into account meaning; together, these applications of GPT vastly widen the set of possible inputs to the task learning system, with minimal upfront engineering cost. 

In Section~\ref{sec:design}, we describe the design of our system in detail and justify our particular application of GPT for the dialog understanding problem. In Section~\ref{sec:eval}, we demonstrate that our system can reproduce the task knowledge acquired by \citet{suddrey} using more complex and ambiguous input language. Finally, in Section~\ref{sec:discussion}, we discuss future expansions of this work, including additional evaluative work and the incorporation of additional instructional modalities.



\section{System Design}
\label{sec:design}

In our system, tasks are conveyed from the \textit{instructor} to the \textit{agent} starting with an initial command provided by the instructor. The command can be an action or a sequence of multiple actions. All actions represented in the command are parsed into predicate-argument structures and matched to corresponding actions already known by the agent. If any actions in a command cannot be matched to semantically similar known actions, they are recursively clarified with new subcommands solicited from the instructor; the recursive clarification of actions into sequences of new actions forms the hierarchical structure of the learned task. One step of this process is depicted in Figure~\ref{fig:arch}, and the full procedure is described in Algorithm~\ref{alg:task_learning}.\footnote{In Algorithm~\ref{alg:task_learning}, the $\leftarrow_{+}$ operator, used on lines 6 and 11, represents concatenation-in-place.}

To enable translation of instructions from the natural dialog space to the semantic action space, our system utilizes a pre-trained GPT model\footnote{We use the \texttt{text-davinci-003} weights for the GPT-3 model, trained by OpenAI.} to perform two separate steps: the extraction of predicate-argument structures from the instructor's utterance (\pgpt), and the mapping of those structures to semantically equivalent actions that have already been acquired (\mgpt).

\begin{algorithm}[H]
\begin{algorithmic}[1]
\caption{Recursive task learning procedure}
\label{alg:task_learning}

\Function{LearnTask}{utterance,known}
\State $T \leftarrow \text{empty tree}$
\State $P \leftarrow \textcolor{blue}{\pgpt}(\text{utterance})$
\For{each $(\text{pred}, \text{args}) \in P$}
\If{$\textcolor{blue}{\mgpt}(\text{pred}) \in \text{known}$}
\State $T \leftarrow_{+}$ ($\textcolor{blue}{\mgpt}(\text{pred})$, args)
\Else
\State \textsc{Print}(``\textit{What does }pred\textit{ mean?}'')
\State $U_{N} \leftarrow$ user input
\State $T_{N} \leftarrow$ \textsc{LearnTask}($U_{N}$, known)
\State known $\leftarrow_{+}$ pred
\EndIf
\EndFor
\State \Return $T$
\EndFunction

\end{algorithmic}
\end{algorithm} 

\subsection{Integration of GPT Models}
\label{sec:use_of_gpt}
We use GPT models to solve atomic, linguistic subroutines within a well-defined algorithm. This allows us to obtain more focused, reliable output than if we had used language models to solve the entire task ``end-to-end''. It also allows us to perform better error correction on a more predictable set of error classes. However, the tasks performed by the language model here are by no means trivial; extraction of predicate-argument structures from text, and the mapping of those structures to known actions, both require resolution of a considerable number of linguistic tasks, some examples of which we will enumerate here, making reference to the example dialog presented in Figure~\ref{fig:val_dialog}:

\begin{enumerate}
    \item \textbf{Syntactic parsing}, a necessary step of semantic parsing in other dialog-based instructional systems, is a challenging task that is highly prone to grammatical and spelling errors, often requiring significant engineering effort to overcome them. \pgpt's virtual mastery of the \textit{form} of natural language, even when that language may be ungrammatical, allows us to avoid most of this engineering effort.
    \item \textbf{Anaphora resolution}, performed by \pgpt, allows constructions like \textit{Pick up the pepper and put it on the counter} to decompose into unambiguous structures like \texttt{pickUp(pepper)} and \texttt{put(pepper, counter)}.
    \item \textbf{Predicate naming}, performed by \pgpt, is important in distinguishing predicates with similar verbs, as verb senses may often include mandatory prepositional arguments, e.g., \texttt{putAway} vs. \texttt{put}.
    \item \textbf{Paraphrase resolution}, performed by \mgpt, is one of the most challenging aspects of unifying natural speech with formal knowledge representations. While other task learning systems struggle with the sheer phrasal variance of natural language, \mgpt~autonomously makes these determinations, such as identifying that both \textit{pull your hand back} and \textit{bring it back to the original position} invoke the known action \texttt{RESETHANDPOSITION}.
\end{enumerate}

In all, our integration of GPT into an otherwise simple algorithm has allowed us to capitalize on the \textit{fluency} of large language models without falling victim to the myriad \textit{failures} that can arise from using them to complete non-linguistic tasks end-to-end---especially recursive tasks and tasks requiring well-defined output.


\section{Evaluation}
\label{sec:eval}

We evaluate our system on the same learned task as \citet{suddrey}, and in the same way, with an author providing dialog input. To highlight the contributions of GPT enumerated in Section~\ref{sec:use_of_gpt}, we provided dialog with ambiguous anaphora, multi-word predicates, and action paraphrasing. The full dialog can be seen in Figure~\ref{fig:val_dialog}. The tasks to be learned are shown in Table~\ref{tab:subtasks}, both with the name of each sub-action in the original evaluation experiment by \citet{suddrey}, and with the name of that action in our system.\footnote{We changed the names of some of these actions to better convey the semantics of each action to GPT; these semantics are important for paraphrase resolution.}

\begin{table}[ht]
    \centering
    \resizebox{\columnwidth}{!}{
    \begin{tabular}{l l l}
    \hline
    \hline
    \textbf{Task Name} & \textbf{Subtasks (orig. names)} & \textbf{Subtasks (our names)} \\
    \hline
        pick\_up & - open & - openHand \\
        & - move & - moveHand \\
        & - close & - closeHand \\
        & - move\_away & - resetHandPosition \\
        \hline
        put & - move & - move \\
        & - open & - openHand \\
        & - move\_away & - resetHandPosition \\
        & - close & - closeHand \\
        \hline
        move & - pick\_up & - pick\_up \\
        & - put & - put \\
        \hline
        put\_away & - move & - move \\
        \hline
        clean & - put\_away & - put\_away
    \end{tabular}
    }
    \caption{The five learned actions for the evaluation method given by \citet{suddrey}, and their subtasks. Both the original authors' names for the subtasks and our modified names are shown.}
    \label{tab:subtasks}
\end{table}

\begin{figure}
    \centering
    \includegraphics[width=\columnwidth]{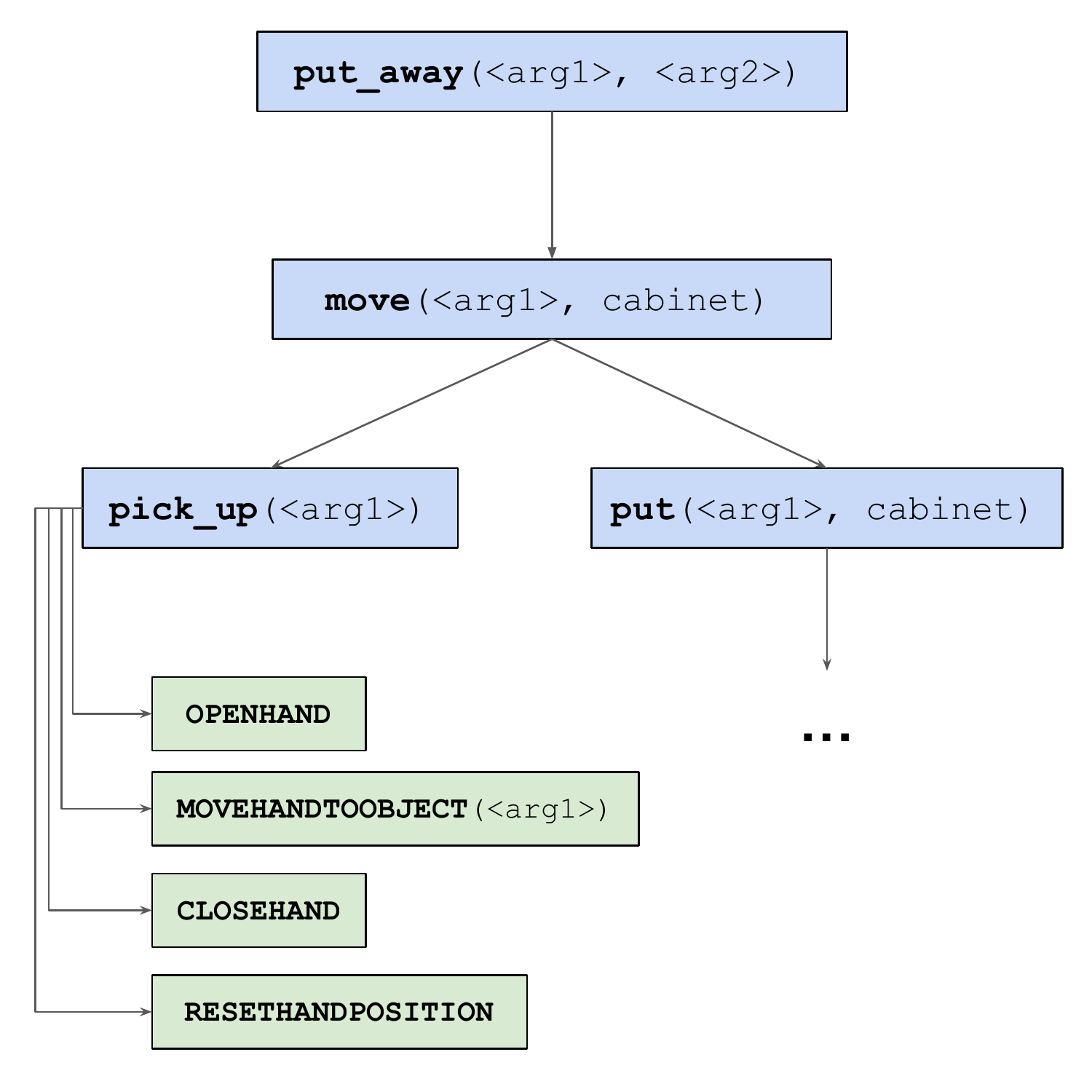}
    \caption{An illustration of a component of the generalized plan learned by our system as a result of the dialog presented in Figure~\ref{fig:val_dialog}.}
    \label{fig:val_plan_tree}
\end{figure}

Our system learned each action from the original work, including their subtask structures, from the provided dialog. The final task decomposition was identical to the structure induced by \citet{suddrey}. The learned structure for the action \texttt{put\_away} is partially presented in Figure~\ref{fig:val_plan_tree}. While further evaluation, especially with grounding in an environment, is called for, we aim here only to show that the task structures themselves can be induced from natural dialog with minimal engineering effort. Future work has the fortunate property of being modular with respect to this induction: environmental symbol grounding, for example, as we discuss in Section~\ref{sec:discussion}, can be implemented as additional subroutines in an extension of Algorithm~\ref{alg:task_learning}.

\section{Discussion and Future Directions}
\label{sec:discussion}

This work is intended to show that the hierarchical task structures learned in the system due to \citet{suddrey} can also be learned using GPT as a means of more fluently handling natural language input. However, the work presented in this short paper is not a complete task learning system: future expansion of this work must incorporate the grounding of objects and plans in an operational environment, as past work has. Additionally, the relatively limited evaluation scheme, chosen here to achieve parity with the reference paper, should be expanded to incorporate a wider range of tasks and dialogs in a future study.

The NTI framework \citep{nti} also highlights many additional instructional modalities, including gestures, images, feedback, and demonstrations. While GPT can be used to amplify the efficacy of linguistic instruction, the integration of these other modalities into a cohesive whole is still an open problem. We believe that the widely predicted upcoming advent of multi-modal transformers, and its combination with symbolic knowledge representations, could allow for interesting forms of joint dialog and image understanding that could be utilized to produce a multi-modal semantic representation for task learning. Exciting times lie ahead in the world of task knowledge representation.


\section{Acknowledgement}
This research was funded in part by Award 2112532 from NSF’s AI-ALOE institute and Awards W911NF2120101 and W911NF2120126 from ARL’s STRONG program. The views, opinions, and findings expressed are the authors’ and should not be taken as representing official views or policies of these funding agencies.


\bibliography{anthology,custom}

\begin{thebibliography}{10}
\expandafter\ifx\csname natexlab\endcsname\relax\def\natexlab#1{#1}\fi

\bibitem[{Allen et~al.(2007)Allen, Chambers, Ferguson, Galescu, Jung, Swift,
  and Taysom}]{plow}
James~F. Allen, Nathanael Chambers, George Ferguson, Lucian Galescu, Hyuckchul
  Jung, Mary~D. Swift, and William Taysom. 2007.
\newblock {PLOW}: A collaborative task learning agent.
\newblock In \emph{AAAI Conference on Artificial Intelligence}.

\bibitem[{Brown et~al.(2020)Brown, Mann, Ryder, Subbiah, Kaplan, Dhariwal,
  Neelakantan, Shyam, Sastry, Askell et~al.}]{gpt-3}
Tom Brown, Benjamin Mann, Nick Ryder, Melanie Subbiah, Jared~D. Kaplan,
  Prafulla Dhariwal, Arvind Neelakantan, Pranav Shyam, Girish Sastry, Amanda
  Askell, et~al. 2020.
\newblock Language models are few-shot learners.
\newblock \emph{Advances in neural information processing systems},
  33:1877--1901.

\bibitem[{Hinrichs and Forbus(2014)}]{forbus}
T.~Hinrichs and K.~Forbus. 2014.
\newblock X goes first: Teaching a simple game through multimodal interaction.
\newblock \emph{Advances in Cognitive Systems}, 3:31--46.

\bibitem[{Huffman and Laird(1993)}]{huffman1993}
Scott~B Huffman and John~E Laird. 1993.
\newblock Learning procedures from interactive natural language instructions.
\newblock In \emph{Proceedings of the 10th International Conference on Machine
  Learning}, pages 143--150.

\bibitem[{Kirk and Laird(2014)}]{rosie}
James~R. Kirk and John~E. Laird. 2014.
\newblock Interactive task learning for simple games.
\newblock \emph{Advances in Cognitive Systems}, 3(13-30):5.

\bibitem[{Laird et~al.(2017)Laird, Gluck, Anderson, Forbus, Jenkins, Lebiere,
  Salvucci, Scheutz, Thomaz, Trafton et~al.}]{itl}
John~E. Laird, Kevin Gluck, John Anderson, Kenneth~D. Forbus, Odest~Chadwicke
  Jenkins, Christian Lebiere, Dario Salvucci, Matthias Scheutz, Andrea Thomaz,
  Greg Trafton, et~al. 2017.
\newblock Interactive task learning.
\newblock \emph{IEEE Intelligent Systems}, 32(4):6--21.

\bibitem[{Laird et~al.(1987)Laird, Newell, and Rosenbloom}]{soar}
John~E. Laird, Allen Newell, and Paul~S. Rosenbloom. 1987.
\newblock Soar: An architecture for general intelligence.
\newblock \emph{Artificial intelligence}, 33(1):1--64.

\bibitem[{MacLellan et~al.(2018)MacLellan, Harpstead, Marinier~III, and
  Koedinger}]{nti}
Christopher~J. MacLellan, Erik Harpstead, Robert~P. Marinier~III, and
  Kenneth~R. Koedinger. 2018.
\newblock A framework for natural cognitive system training interactions.
\newblock \emph{Advances in Cognitive Systems}, 6:1--16.

\bibitem[{Suddrey et~al.(2017)Suddrey, Lehnert, Eich, Maire, and
  Roberts}]{suddrey}
Gavin Suddrey, Christopher Lehnert, Markus Eich, Frederic Maire, and Jonathan
  Roberts. 2017.
\newblock \href {https://doi.org/10.1109/LRA.2016.2588584} {Teaching robots
  generalizable hierarchical tasks through natural language instruction}.
\newblock \emph{IEEE Robotics and Automation Letters}, 2(1):201--208.

\bibitem[{Tenney et~al.(2019)Tenney, Das, and Pavlick}]{bert-rediscovers-nlp}
Ian Tenney, Dipanjan Das, and Ellie Pavlick. 2019.
\newblock \href {https://doi.org/10.18653/v1/P19-1452} {{BERT} rediscovers the
  classical {NLP} pipeline}.
\newblock In \emph{Proceedings of the 57th Annual Meeting of the Association
  for Computational Linguistics}, pages 4593--4601, Florence, Italy.
  Association for Computational Linguistics.

\end{thebibliography}
\bibliographystyle{acl_natbib}




\end{document}